\documentclass[aps,prb,reprint,superscriptaddress,floatfix,longbibliography]{revtex4-1}
\usepackage{graphicx,float}
\usepackage{amsmath,amssymb,mathrsfs,dcolumn}
\usepackage{gensymb}
\usepackage[usenames]{color}
\usepackage{subfigure}
\usepackage[normalem]{ulem}
\usepackage{comment}
\usepackage{hyperref}
\hypersetup{colorlinks=true, linkcolor=blue, citecolor=red, urlcolor=blue}

\begin{document}
%\title{Relaxation of squeezed magnons}
\title{Master equation approach to magnon relaxation and dephasing}
\author{H. Y. Yuan}
\author{W. P. Sterk}
\affiliation{Institute for Theoretical Physics, Utrecht University, 3584CC Utrecht, The Netherlands}
\author{Akashdeep Kamra}
\affiliation{Condensed Matter Physics  Center (IFIMAC) and Departamento de F\'{i}sica Te\'{o}rica de la Materia Condensada, Universidad Aut\'{o}noma de Madrid, E-28049 Madrid, Spain}
\author{Rembert A. Duine}
\affiliation{Institute for Theoretical Physics, Utrecht University, 3584CC Utrecht, The Netherlands}
\affiliation{Department of Applied Physics, Eindhoven University of Technology, P.O. Box 513, 5600 MB Eindhoven, The Netherlands}
\date{\today}
\date{\today}

\begin{abstract}
There has been a recent upsurge of interest in the quantum properties of magnons for quantum information processing. An important issue is to examine the stability of quantum states of magnons against various relaxation and dephasing channels. Since the interaction of magnons in magnetic systems may fall in the ultra-strong and even deep-strong coupling regimes, the relaxation process of magnon states is quite different from the more common quantum optical systems. Here we study the relaxation and dephasing of magnons based on the Lindblad formalism and derive a generalized master equation that describes the quantum dynamics of magnons. Employing this master equation, we identify two distinct dissipation channels for squeezed magnons, i.e., the local dissipation and collective dissipation, which play a role for both ferromagnets and antiferromagnets. The local dissipation is caused by the independent exchange of angular momentum between the magnonic system and the environment, while the collective dissipation is dressed by the parametric interactions of magnons and it enhances the quantumness and thermal stability of squeezed magnons. Further, we show how this formalism can be applied to study the pure dephasing of magnons caused by four-magnon scattering and magnon-phonon interactions. Our results provide the theoretical tools to study the decoherence of magnons within a full quantum-mechanical framework and further benefit the use of quantum states of magnons for information processing.
\end{abstract}

\maketitle

\section{Introduction}
Magnons are the collective excitation of ordered magnets and they can carry information for data storage and processing. Because of the low dissipation rate and long spin diffusion length of magnons in magnetic insulators, the information coded in magnons can, in principle, weaken Joule heating problems and thus it is a desirable candidate for durable information processing \cite{ChumakNP2015}. As bosonic particles, magnons share considerable similarities with photons and the quantum states of magnons such as single-magnon state \cite{YuanPRB2020,Lachance-QuirionSciAdv2017}, squeezed states \cite{ZhaoPRL2004,KamraPRL2016,LiPRA2019,KamraAPL2020,YuanPRB2021B}, Schr\"{o}dinger cat states \cite{SharmaPRB2021,SunPRL2021} and quantum many-body states \cite{Yuanreview2022} have already been proposed in magnonic systems for information processing. The emerging field, dubbed ``\textit{quantum magnonics}" \cite{Yuanreview2022}, has started to attract significant interest due to its potential to push the traditional studies of magnonics to the quantum limit, and for its convenient docking with mature solid-state quantum platforms, such as cavity photons, mechanical oscillators, superconducting qubits and single-spin qubits.

To achieve desirable information processing using hybrid magnonic systems, a crucial objective is to maintain the coherence of magnons for a longer time. The decoherence in a magnetic system may come from the coupling of magnons with conduction electrons, lattice vibrations or phonons, impurities and magnon-magnon scattering process. The standard approach to characterize the dissipation of magnons in the past century is to introduce a phenomenological parameter called the Gilbert damping in the classical magnetization dynamics \cite{Landau1935,Gilbert2004}. The linewidth of the magnetic resonance under external microwave driving is proportional to the Gilbert damping. By analyzing the frequency and temperature dependence of the linewidth, one may extract the various contributions to the damping \cite{AriasPRB1999, McMPRL2003, NembachPRL2013,KhodPRL2020}. In the framework of quantum optics, one usually introduces a decay rate of the photon mode, together with a stochastic field acting on the photon in the Heisenberg-Langevin equation, according to the fluctuation-dissipation theorem \cite{DFWallsbook}. However,
different from photons, magnons in the magnets are naturally coupled with each other through either anisotropy or exchange interactions \cite{Whitebook}. The coupling strength can easily reach ultra-strong and even deep-strong coupling regimes \cite{Yuanreview2022}. For example, in a two-sublattice antiferromagnet (AFM), the coupling of sublattice magnons is on the order of exchange interaction, which is as strong as the on-site frequency of the magnons. One has to consistently include the interactions among magnons to get a reasonable prediction of both the dynamic and equilibrium properties of the magnonic systems, especially the decoherence time of magnons. It has been known that the strong interactions inside a hybrid quantum optical system can cause additional dissipation channels and modify the dynamic equations \cite{CarmJPA1973,RivasNJP2010,BanIJTP2017,BetQST2020}.  For magnonic systems, how the strong interaction among magnons influences the relaxation processes channels, especially its influence on the robustness of the quantum properties of magnons, has not been examined so far.

On the other hand, up till now, most of the studies on the magnon decoherence only account for relaxation process, while how pure dephasing without relaxation process almost gains no attention. If one intends to use quantum states of magnons for quantum information processing, the pure dephasing channel should be treated equal footing as the relaxation channel and be addressed properly, resembling the situation in more conventional qubit systems \cite{YangPRL2008,BoissPRA2009,QTbook,Nielsenbook}.

In this article, we study the relaxation and dephasing of magnons in both ferromagnets and antiferromagnets based on the Lindblad formalism. We derive a generalized master equation for the dynamics of magnons under the Born-Markov approximation. In particular, we identify the interaction induced collective relaxation channels of magnons. This channel could not only enhance the entanglement of magnons, but also makes the quantum correlations of magnons more stable against thermal fluctuations, which is very different from the local dissipation that will destroy the quantumness of the systems. Second, we show that the antiferromagnetic magnons are subject to a decoherence-free evolution in ideal conditions when they are subject to a common bath. The comparison with the relaxation process within classical equation of motion is discussed in detail. Further, we show how our theoretical formalism can recover the dephaisng channels of magnons through four-magnon interaction and magnon-phonon coupling.

Our work is organized as follows: In Sec. \ref{general_me}, we present the generalized formalism to consider the magnon relaxation caused by their interaction with the environment. In Sec. \ref{sec_single_mode}, we apply the general formalism to study the stability of the single-mode squeezed state in an anisotropic ferromagnet. Then we introduce the two-mode squeezed magnons \cite{YuanPRB2020B,KamraPRB2019} in a two-sublattice AFM and their relaxation channel subject to both two separate and one common bath and show how decoherence-free evolution of magnons can be realized in Sec. \ref{sec_two_mode}. The classical-quantum analogue of the dissipation channels is also discussed in detail. In Sec. \ref{sec_dephase}, we apply the generalized master equation to address the dephasing of magnons caused by magnon-magnon interaction and magnon-phonon scattering. In Sec. \ref{sec_conclusion}, we come to the discussions and conclusions. Unless stated otherwise, the reduced Planck constant $\hbar$ is set to one.

\section{Master equation approach} \label{general_me}
Let us consider a magnonic system described by the Hamiltonian $\hat{\mathcal{H}}_\mathrm{FM}$, which is expressed in terms of the magnon creation ($\hat{a}^\dagger$) and annihilation ($\hat{a}$) operators in general.  To study the decoherence of the magnon state against thermal fluctuation, we model the environment as a collection of harmonic oscillators, i.e., $\hat{\mathcal{H}}_\mathrm{R}=\sum_i \omega_i \hat{c}_i^\dagger \hat{c}_i$, with commutation relations $[\hat{c}_i,\hat{c}_j^\dagger]=\delta_{ij}$. The interaction of magnons with the bath is through the coherent transfer between magnons and bath, $\hat{\mathcal{H}}_\mathrm{int}=\sum_i g_i ( \hat{a}\hat{c}_i^\dagger +  \hat{a}^\dagger \hat{c}_i)$.
Note that the bath modes in the environment may include phonons, electron-hole pairs, or impurities. Here we do not specify them and only assume a phenomenological coupling coefficient $g_i$ for simplicity. A more detailed consideration of the various baths will not influence the essential physics presented below. It will, however, determine the temperature and frequency dependence of the coupling coefficients \cite{MichaelJAP2002,SafonovJAP2003,AndreasPRB2014}. With these assumptions, the total Hamiltonian of the system reads
\begin{equation}
\hat{\mathcal{H}}_\mathrm{T}=\hat{\mathcal{H}}_\mathrm{FM} + \hat{\mathcal{H}}_\mathrm{R} + \hat{\mathcal{H}}_\mathrm{int}.
\end{equation}

Since the total system is Hermitian, its dynamic evolution should follow the master equation $\partial \hat{\rho}_T(t) / \partial t = i[\hat{\rho}_T(t), \hat{\mathcal{H}}_\mathrm{T}]$, where $\hat{\rho}_T$ is the density matrix of the total system. To eliminate the non-interacting terms in the total Hamiltonian $\hat{\mathcal{H}}_\mathrm{T}$, we transfer to the interaction picture and trace out the degree of freedoms for the bath to derive the evolution of the ferromagnetic magnons as
\begin{equation}\label{master_eq1}
\frac{d \tilde{\rho}(t)}{dt}=-\int_0^{t} dt' tr_R \left [ \tilde{\mathcal{H}}_\mathrm{int}(t), [\tilde{\mathcal{H}}_\mathrm{int}(t'),\tilde{\rho}_T(t')]\right ],
\end{equation}
where $\tilde{\rho}(t) \equiv tr_R\tilde{\rho}_\mathrm{T}(t)$. Here all the operators in the interaction picture are labeled with an overhead tilde. Under the Born-Markov approximation, the master equation \eqref{master_eq1} can be reduced to \cite{LindbladCMP1976,GoriniJMP1976,Carmbook1991,ManzanoAIP2020}
\begin{equation}\label{master_eq2}
\frac{d \tilde{\rho}(t)}{dt}=-\int_0^{t} dt' tr_R \left [ \tilde{\mathcal{H}}_\mathrm{int}(t), [\tilde{\mathcal{H}}_\mathrm{int}(t'),\tilde{\rho}(t)\otimes \tilde{\rho}_R(0)]\right ],
\end{equation}
where $\tilde{\rho}_R(0)$ is the initial density matrix of the bath and is assumed to be independent of time due to its large degrees of freedoms. Now the state of magnons $\tilde{\rho}(t)$ does not explicitly depend on its history $\tilde{\rho}(t')$ ($t'<t$). Following the standard notations to rewrite the interaction in the form $\hat{\mathcal{H}}_\mathrm{int}=\sum_i \hat{s}_i \otimes \hat{\Gamma}_i$ \cite{ManzanoAIP2020}, where $\hat{s}_i$ and $\hat{\Gamma}_i$ are respectively the operators acting in the spaces of magnonic system and bath, the master equation \eqref{master_eq2} can be finally written as
\begin{equation}\label{gme}
\begin{aligned}
\frac{d \tilde{\rho}(t)}{dt}&=-\sum_{i,j} [\tilde{s}_i(t) \tilde{s}_j(t') \tilde{\rho}(t)-\tilde{s}_j(t') \tilde{\rho}(t) \tilde{s}_i(t) ]\langle \tilde{\Gamma}_i(t) \tilde{\Gamma}_j(t')\rangle \\
&-\sum_{i,j} [\tilde{\rho}(t) \tilde{s}_j(t') \tilde{s}_i(t)-\tilde{s}_i(t) \tilde{\rho}(t) \tilde{s}_j(t') ]\langle \tilde{\Gamma}_j(t') \tilde{\Gamma}_i(t)\rangle.
\end{aligned}
\end{equation}
This equation is very general and has been widely used to study the dynamic behavior of open quantum system in quantum optics \cite{LindbladCMP1976,GoriniJMP1976,Carmbook1991,ManzanoAIP2020}. Below we shall study the relaxation of both the single-mode and two-mode squeezed magnons and pure dephasing of magnons based on this master equation \cite{YuanArxiv2022}.

\section{single-mode squeezed magnon}\label{sec_single_mode}
Let us consider a biaxial ferromagnet described by the Hamiltonian
\begin{equation}\label{Ham_FM}
\hat{\mathcal{H}}_{\mathrm{FM}}=-J\sum_{\langle ij\rangle} \hat{\mathbf{S}}_i\cdot \hat{\mathbf{S}}_j - \sum_j \left [ K_z(\hat{S}_j^z)^2 - K_x (\hat{S}_j^x)^2+H \hat{S}_j^z \right ],
\end{equation}
where $\hat{\mathbf{S}}_i$ is the spin operator on the $i$th site with spin number $S$, $J>0$ is the ferromagnetic exchange interaction between neighboring spins, $K_z>0$ and $K_x>0$ are respectively the easy-axis and hard-axis anisotropies, and $H$ is an external field along the $z-$axis. The classical ground state of the system is $\langle \mathbf{S}_i \rangle=Se_z$. To consider the magnon excitation above this ground state, we perform Holstein-Primakoff (HP) transformation of the spin operators \cite{HolsteinPRB1940}, i.e., $\hat{S}_{iz}=S-\hat{a}^\dagger \hat{a}$, $\hat{S}^+_i=\sqrt{2S-\hat{a}^\dagger \hat{a}} \hat{a}$, $\hat{S}^-_i=\hat{a}^\dagger \sqrt{2S-\hat{a}^\dagger \hat{a}}$, and substitute them into \eqref{Ham_FM} to derive the effective Hamiltonian of magnon excitations up to the quadratic order as
\begin{equation}
\hat{\mathcal{H}}_\mathrm{FM}=\omega_a \hat{a}^\dagger \hat{a} + g_{aa}(\hat{a}^\dagger \hat{a}^\dagger + \hat{a}\hat{a}),
\label{Hamk}
\end{equation}
where $\omega_a=2K_zS+K_xS+H$, $g_{aa}=K_xS/2$. Here we considered uniform precession mode of spins such that magnon operator $\hat{a}$ is spatial independent and the exchange coefficient $J$ does not appear in the Hamiltonian. Nevertheless, the formalism presented below is straightforward to be generalized to study the evolution of the magnon modes with non-zero wavevectors.

The ground state of the magnonic system described by Hamiltonian \eqref{Hamk} is a single-mode squeezed state $|GS\rangle =\mathrm{exp}(r/2(\hat{a}\hat{a} - \hat{a}^\dagger \hat{a}^\dagger ))|0\rangle$ with squeezing parameter $r=\mathrm{arctanh}(2g_{aa}/\omega_a)/2>0$ and magnon occupation $\langle \hat{a}^\dagger \hat{a}\rangle = \sinh^2r$. The uncertainties of the total spin components $\hat{S}_x=\sum_i \hat{S}_i^x$ and $\hat{S}_y=\sum_i \hat{S}_i^y$ in this ground state are respectively, $\Delta S_x \equiv\sqrt{\langle \hat{S}_x^2\rangle -\langle \hat{S}_x \rangle ^2}= e^{-r}\sqrt{NS/2}$ and $\Delta S_y \equiv \sqrt{\langle S\hat{}_y^2 \rangle-\langle \hat{S}_y \rangle ^2}=e^{r}\sqrt{NS/2}$. It is clear that the noise of $\hat{S}_x$ is squeezed below the symmetric limit at the sacrifice of uncertainty increase in the $\hat{S}_y$ component. The Heisenberg uncertainty relation is still maintained.

Now we are ready to apply the general formalism to study the quantum decoherence of squeezed magnons. The magnon and bath operators in this special case respectively read
\begin{widetext}
\begin{subequations}
\begin{align}
&\tilde{s}_1(t)= \left [\frac{\omega_r - \omega_a}{2\omega_r}  \hat{a} - \frac{g_{aa}}{\omega_r}\hat{a}^\dagger\right ] e^{i\omega_r t} + \left [ \frac{\omega_r + \omega_a}{2\omega_r}  \hat{a} + \frac{g_{aa}}{\omega_r}\hat{a}^\dagger\right ] e^{-i\omega_r t}, \tilde{\Gamma}_1(t)= \sum_i \kappa_i \hat{c}_i^\dagger e^{i\omega_r t},\\
&\tilde{s}_2(t)= \left [\frac{\omega_r - \omega_a}{2\omega_r}  \hat{a}^\dagger - \frac{g_{aa}}{\omega_r}\hat{a}\right ] e^{-i\omega_r t} + \left [ \frac{\omega_r + \omega_a}{2\omega_r}  \hat{a}^\dagger + \frac{g_{aa}}{\omega_r}\hat{a}^\dagger\right ] e^{i\omega_r t},\tilde{\Gamma}_2(t)= \sum_i \kappa_i \hat{c}_i e^{-i\omega_r t},
\end{align}
\end{subequations}
\end{widetext}
where $\omega_r=\sqrt{\omega_a^2 - 4 g_{aa}^2}$ is the resonance frequency of the magnetic system. By substituting all these variables into the master equation (\ref{gme}), simplifying the integrals using bath correlations at equilibrium conditions, and finally transferring it back to the original Schr\"{o}dinger picture, we have
\begin{widetext}
\begin{equation}\label{singlemode_me}
\begin{aligned}
\frac{d\hat{\rho}}{dt}=i[\hat{\rho},\hat{\mathcal{H}}_{\mathrm{FM}}]
+ \kappa (n_{\mathrm{th}}+1) \left[\eta_{\omega\omega} \mathcal{L}_{aa} + \eta_{\omega g}(\mathcal{L}_{aa^\dagger} + \mathcal{L}_{a^\dagger a})+\eta_{gg}\mathcal{L}_{a^\dagger a^\dagger} \right ][\hat{\rho}]\\
+ \kappa n_{\mathrm{th}}\left[\eta_{\omega\omega} \mathcal{L}_{a^\dagger a^\dagger} + \eta_{\omega g}(\mathcal{L}_{aa^\dagger} + \mathcal{L}_{a^\dagger a}) + \eta_{gg}\mathcal{L}_{a a}\right ][\hat{\rho}],
\end{aligned}
\end{equation}
\end{widetext}
where $\mathcal{L}_{AB}[\hat{\rho}]=2\hat{A}\hat{\rho} \hat{B}^\dagger - \hat{\rho} \hat{A}^\dagger \hat{B} - \hat{A}^\dagger \hat{B} \hat{\rho}$ is the generalized Lindblad operator, and $n_{\mathrm{th}} = (e^{\omega_r/k_BT} -1)^{-1}$ is the Bose-Einstein distribution function with $T$ being the bath temperature, $\kappa$ is the effective relaxation rate defined as $\kappa \equiv g(\omega_r) D(\omega_r)$, where $D(\omega_r)$ is the total density states of bath modes at the resonance frequency $\omega_r$.

\begin{figure}
  \centering
  \includegraphics[width=0.45\textwidth]{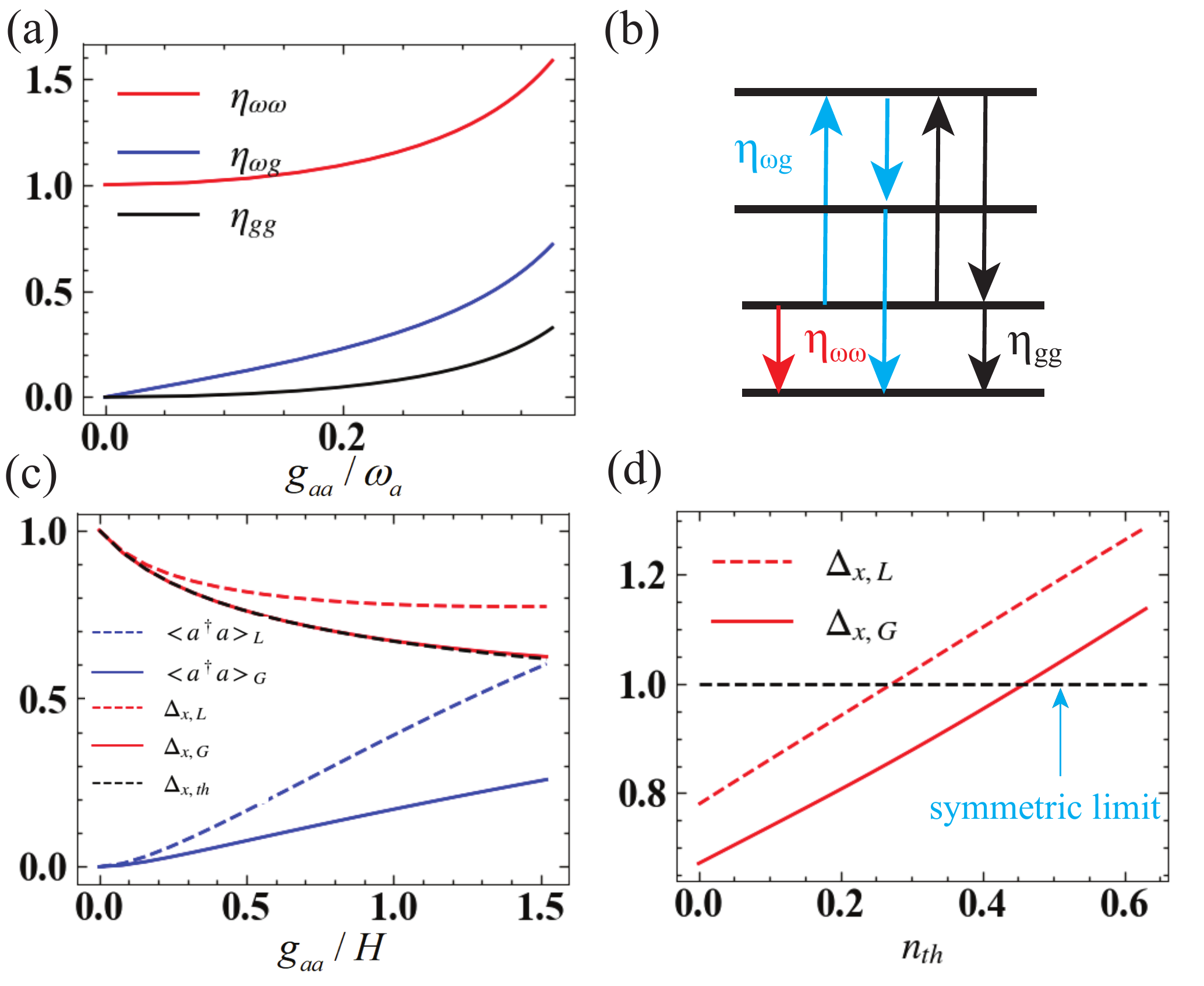}\\
  \caption{(a) Local ($\eta_{\omega \omega}$) and collective ($\eta_{\omega g},\eta_{g g}$) relaxation coefficients as a function of the anisotropy strength $g_{aa}$. $H=1.0$ T, $K_z=0$. (b) Schematic picture of the three types of relaxation process. (c) Magnon occupation and uncertainty of $\hat{S}_x$ as a function of anisotropy at zero temperature ($n_{\mathrm{th}}=0$). The subscript label $L$ means only the local dissipation $\eta_{\omega \omega}$ is included, while $G$ means both local and collective dissipation channels are taken into account. The expected value of the uncertainty in the ground state is plotted as black dashed lines. (d) Uncertainty of $\hat{S}_x$ as a function of thermal occupation $n_{\mathrm{th}}$. $g_{aa}/H=1.0$. The horizontal dashed line represents the symmetric limit. Above the horizontal line, both the uncertainties of $\hat{S}_x$ and $\hat{S}_y$ are larger than the symmetric limit, even though the uncertainty of $\hat{S}_x$ is still smaller than $\hat{S}_y$.}\label{fig1}
\end{figure}

The three distinguishable relaxation coefficients $\eta_{\omega \omega},\eta_{\omega g},\eta_{gg}$ are respectively defined as
\begin{equation}
\eta_{\omega \omega} = \frac{(\omega_r+\omega_a)^2}{4\omega_r^2},
\eta_{\omega g} = \frac{ g_{aa}(\omega_r+\omega_a)}{2\omega_r^2},
\eta_{gg} = \frac{ g_{aa}^2}{\omega_r^2}.
\end{equation}
Here $\eta_{\omega \omega}$ is the leading-order relaxation rate, and $\eta_{\omega g}$ and $\eta_{g g}$ can respectively be viewed as the first and second order corrections of the dissipation channels caused by the squeezing interactions. Their dependencies on the hard-axis anisotropy of the system $g_{aa}$ (or equivalently $K_x$) are shown in Fig. \ref{fig1}(a). As the anisotropy parameter $g_{aa}$ increases, both $\eta_{\omega g}$ and $\eta_{g g}$ becomes comparable with the leading-order terms $\eta_{\omega \omega}$ and thus they can not be neglected any longer. Note that we can derive the dynamic equation \eqref{singlemode_me} alternatively, by first transferring to the diagonal basis of Hamiltonian \eqref{Hamk} through a Bogliubov transformation, deriving the master equation in this basis and then transfer back to the original basis.
Equation \eqref{singlemode_me} is a powerful result that allows us to study the quantum dynamics of squeezed magnons starting from any initial state, as well as the equilibrium properties of the squeezed magnons.

An intuitive picture to understand these corrections of dissipation channels is illustrated in Fig. \ref{fig1}(b). Without squeezing ($g_{aa}=0,\eta_{\omega \omega}=1, \eta_{\omega g}=\eta_{gg}=0$), magnons can only directly relax to the ground state at zero temperature. The resulting master equation is reduced to the form
\begin{equation}
\frac{d\hat{\rho}}{dt}=i[\hat{\rho},\hat{\mathcal{H}}_{\mathrm{FM}}]
+ \kappa (n_{\mathrm{th}}+1) \mathcal{L}_{aa}[\hat{\rho}] +\kappa n_{\mathrm{th}} \mathcal{L}_{a^\dagger a^\dagger}[\hat{\rho}].
\end{equation}
With squeezing, the magnons may first be excited to higher energy levels ($g_{aa} \hat{a}^\dagger \hat{a}^\dagger$) and then relax, which gives the first-order correction ($\eta_{\omega g}$). Further, the magnons
on energy level $| n \rangle$ may be excited to $| n +2\rangle$ and fall back to $| n \rangle$ and then relax, this correspond to the second-order correction ($\eta_{g g}$). These higher-order processes involve the simultaneous excitation/annihilation of two or more magnons, hence we refer to them as collective relaxation channels.

%$\kappa=0.04$ is used to reach the steady state

To study how the collective relaxation channel influences the quantumness of the system, we simulate the quantum master equation (\ref{singlemode_me}) and show the uncertainty of the $S_x$ component as a function of anisotropy in Fig. \ref{fig1}(c). Here the uncertainty is defined as $\Delta_x = \sqrt{2/NS} \Delta S_x $. As the anisotropy parameter $g_{aa}$ increases, the generalized master equation recovers a steady state with much stronger squeezing ($\Delta_{x,G}$, red solid line) than the result by only including the local dissipations ($\Delta_{x,L}$, red dashed lines). By zooming in the steady state, we found it is exactly the ground state of the system $|GS\rangle$, and thus they share the same magnon occupation and squeezing of $\hat{S}_x$ ($\Delta_{x,th}$, black dashed lines). As a comparison, the system cannot relax to ground state if only the local dissipation is considered ($\Delta_{x,L}$, red dashed lines). Therefore, strictly speaking, the generalized master equation is essential to recover the correct equilibrium properties of the interacting magnon system.

Figure \ref{fig1}(d) shows the influence of temperature on the magnon squeezing. Firstly, as the temperature increases, the uncertainty $\Delta_{x,G}$ gradually increases above the symmetric limit. The critical value of the temperature is around 2 K for $n_{\mathrm{th}}=0.46$ and $\omega_r/2\pi = 48.5$ GHz. This behavior clearly shows that magnon squeezing is a low temperature quantum phenomenon. It is different from the elliptical trajectory of spin precession under the influence of anisotropy, which always exists in anisotropic systems regardless of temperature \cite{YuanPRB2021}. Secondly, when only the local dissipation channel is taken into account, the squeezing will die at $n_{\mathrm{th}}=0.27$ smaller than the value of 0.46 when both local and collective dissipation channels are taken into account. This comparison suggests that the collective relaxation channel can increase the thermal stability of the quantumness of magnons.

\section{Two-mode squeezed magnon}\label{sec_two_mode}
Now we study the relaxation of two-mode squeezed magnons. Two-sublattice AFM hosts two types of magnons excited on each sublattice and it provides a natural platform to study the quantum correlations of magnons. The ground-state properties of the system guarantees that the generation of magnons in one sublattice is always accompanied by the generation of another magnon in the other sublattice, i.e., the interaction of the sublattice magnons is parametric-type \cite{Yuanreview2022}. Therefore, they naturally form a two-mode squeezed state \cite{YuanPRB2021,YuanPRB2020B,KamraPRB2019,KamraAPL2020}. Furthermore, the sublattice magnons are coupled through exchange interaction, which is comparable with the on-site frequencies of magnons and thus reaches the deep-strong coupling regime. We will see how such a strong interaction influences the relaxation channel of the system.

We consider a two-sublattice AFM described by the Hamiltonian
\begin{equation}
\hat{\mathcal{H}}_{\mathrm{AFM}}=J \sum_{j,\delta}  \hat{\mathbf{S}}_j \cdot \hat{\mathbf{S}}_{j+\delta}
- K_z \sum_j (\hat{S}_j^z)^2- \sum_j \mathbf{H} \cdot \hat{\mathbf{S}}_{j},
\end{equation}
where $J>0$ is the antiferromagnetic exchange coupling between neighboring spins, $K_z > 0$ is the easy-axis anisotropy.
Given the external field applied along the $z-$axis ($\mathbf{H}=He_z$), the classical ground state of the systems when the field $H$ is below the spin-flop transition is a N\'{e}el state, i.e., $\langle \hat{\mathbf{S}}_{2l} \rangle =Se_z, \langle \hat{\mathbf{S}}_{2l+1} \rangle=-Se_z$. Then we can follow the standard HP transformation to quantize the magnon excitation above this ground state as
\begin{equation}
\hat{\mathcal{H}}_{\mathrm{AFM}}=  \omega_a \hat{a}^\dagger \hat{a} + \omega_b \hat{b}^\dagger \hat{b} +g_{ab}(\hat{a}^\dagger
\hat{b}^\dagger + \hat{a} \hat{b}),
\end{equation}
where $\omega_{a,b}=H_{\mathrm{ex}} + H_{\mathrm{an}} \pm H$, $H_{\mathrm{ex}}=2ZJS^2, H_{\mathrm{an}}=2K_zS$ with $Z$ being the coordination number, $g_{\mathrm{ab}}=H_\mathrm{ex}$ is the coupling of magnon modes excited on the two sublattices.
Again, we focus on the uniform precession mode. The ground-state entanglement of magnons taking the contribution of finite wavevectors into account can be found in Refs. \cite{HartmannPRB2021,MousolouPRB2021,WuhrerPRB2021}

To describe the interaction of magnons with the environment, we have to clarify whether the two types of magnons are subject to two separate bath or one common bath. Both cases contain local and collective dissipation channels, but they will result in very different relaxation process as we shall discuss below.

\subsection{Two separate bath} \label{sec_afm_twobath}

When each sublattice magnon is coupled to their own bath, the total Hamiltonian of the hybrid system reads
$\hat{\mathcal{H}}_\mathrm{T}=\hat{\mathcal{H}}_\mathrm{AFM} + \hat{\mathcal{H}}_\mathrm{R} + \hat{\mathcal{H}}_\mathrm{int}$, where
\begin{subequations}
\begin{align}
&\hat{\mathcal{H}}_\mathrm{R} = \sum_i \omega_i^{(c)} \hat{c}_i^\dagger \hat{c}_i + \sum_i \omega_i^{(d)} \hat{d}_i^\dagger \hat{d}_i, \\
&\hat{\mathcal{H}}_\mathrm{int} = g_i\sum_i ( \hat{a}\hat{c}_i^\dagger +  \hat{a}^\dagger \hat{c}_i) + g_i \sum_i ( \hat{b}\hat{d}_i^\dagger + \hat{b}^\dagger \hat{d}_i).
\end{align}
\label{Hafmk}
\end{subequations}
Since the two baths are independent, the bath operators should commute with each other, i.e.,
\begin{equation}\label{afm_commutator}
[\hat{c}_i,\hat{d}_j]=[\hat{c}_i^\dagger,\hat{d}_j^\dagger]=[\hat{c}_i,\hat{d}_j^\dagger] = [\hat{c}_i^\dagger,\hat{d}_j]=0.
\end{equation}
In the interaction picture, we have

\begin{figure}
  \centering
  \includegraphics[width=0.45\textwidth]{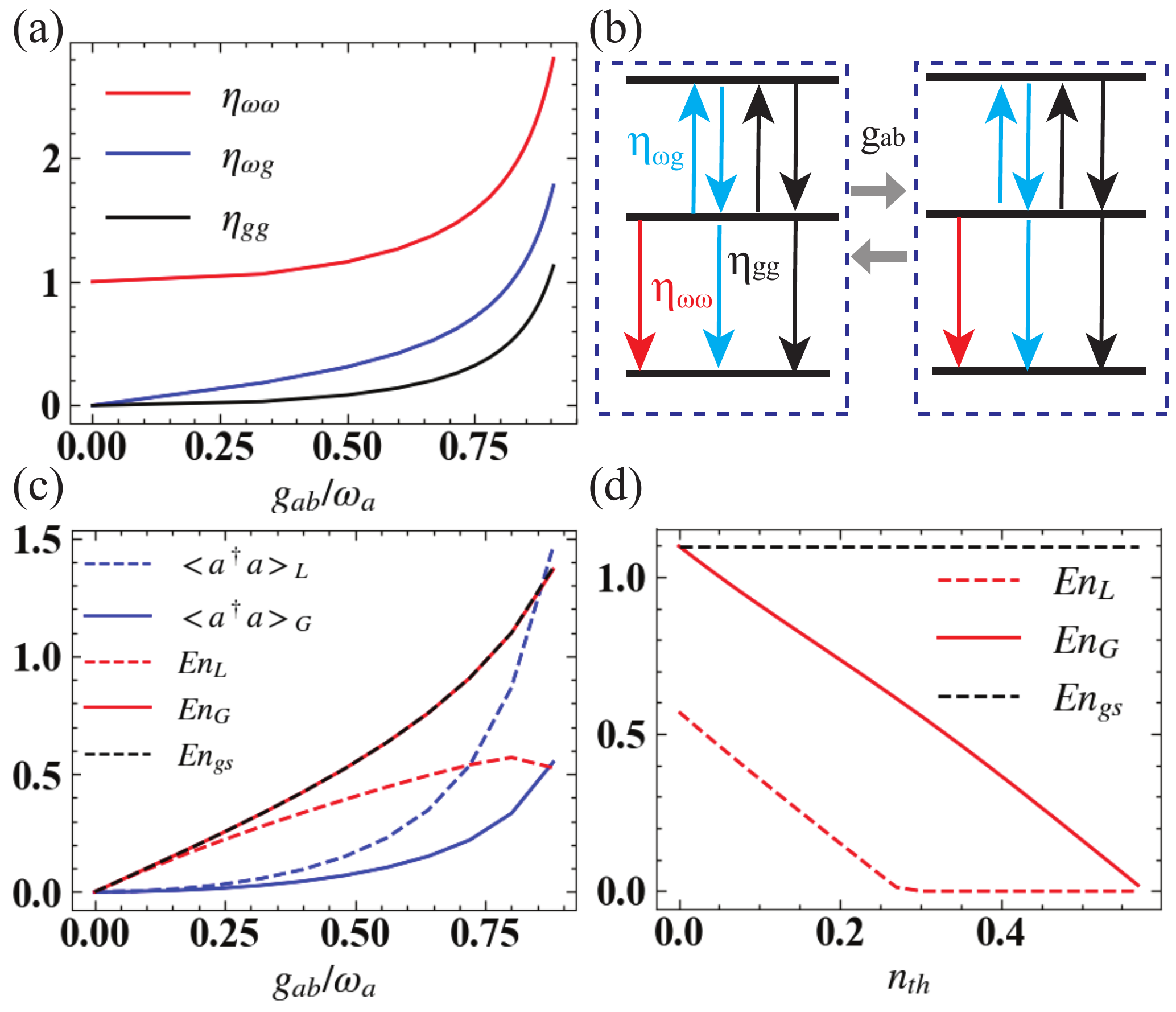}\\
  \caption{(a) Local ($\eta_{\omega \omega}$) and collective ($\eta_{\omega g},\eta_{g g}$) dissipation coefficients as a function of the anisotropy. $H=0$, $H_{an}=1$ T. (b) Schematic picture of the three dissipation processes. (c) Magnon occupation and entanglement of sublattice magnons as a function of anisotropy at zero temperature ($n_{\mathrm{th}}=0$). The subscript label L means only the local dissipation is included while G means both local and nonlocal dissipation channels are taken into account. The expected value of the magnon-magnon entanglement in the ground state is plotted as black dashed lines. (d) Entanglement of the sublattice magnons as a function of thermal occupation. $g_{ab}/\omega_a=0.8,\kappa=0.02$.}\label{fig2}
\end{figure}

\begin{widetext}
\begin{subequations}\label{operator_new_picture}
\begin{align}
&\tilde{s}_1(t)= \left [\frac{\omega_r + \omega_a}{2\omega_r}  \hat{a}^\dagger + \frac{g_{ab}}{2\omega_r}\hat{b}\right ] e^{i\omega_r t} + \left [ \frac{\omega_r - \omega_a}{2\omega_r}  \hat{a}^\dagger - \frac{g_{ab}}{\omega_r}\hat{b}\right ] e^{-i\omega_r t}, \tilde{\Gamma}_1(t)= \sum_i \kappa_i \hat{c}_i^\dagger e^{i\omega_r t},\\
&\tilde{s}_2(t)= \left [\frac{\omega_r + \omega_a}{2\omega_r}  \hat{a} + \frac{g_{ab}}{2\omega_r}\hat{b}^\dagger\right ] e^{-i\omega_r t} + \left [ \frac{\omega_r - \omega_a}{2\omega_r}  \hat{a} - \frac{g_{ab}}{\omega_r}\hat{b}^\dagger \right ] e^{i\omega_r t}, \tilde{\Gamma}_2(t)= \sum_i \kappa_i \hat{c}_i e^{-i\omega_r t},\\
&\tilde{s}_3(t)= \left [\frac{\omega_r + \omega_a}{2\omega_r}  \hat{b}^\dagger + \frac{g_{ab}}{2\omega_r}\hat{a}\right ] e^{i\omega_r t} + \left [ \frac{\omega_r - \omega_a}{2\omega_r}  \hat{b}^\dagger - \frac{g_{ab}}{\omega_r}\hat{a}\right ] e^{-i\omega_r t}, \tilde{\Gamma}_3(t)= \sum_i \kappa_i \hat{d}_i^\dagger e^{i\omega_r t},\\
&\tilde{s}_4(t)= \left [\frac{\omega_r + \omega_a}{2\omega_r}  \hat{b} + \frac{g_{ab}}{2\omega_r}\hat{a}^\dagger\right ] e^{-i\omega_r t} + \left [ \frac{\omega_r - \omega_a}{2\omega_r}  \hat{b} - \frac{g_{ab}}{\omega_r}\hat{a}^\dagger \right ] e^{i\omega_r t}, \tilde{\Gamma}_4(t)= \sum_i \kappa_i \hat{d}_i e^{-i\omega_r t},
\end{align}
\end{subequations}
\end{widetext}
where we disregard the applied magnetic field for simplicity, such that $\omega_a=\omega_b$, and the antiferromagnetic resonance frequency $\omega_r=\sqrt{\omega_a^2-g_{ab}^2}$.
By employing the commutators \eqref{afm_commutator}, the cross correlation terms $\langle \tilde{\Gamma}_i(t) \tilde{\Gamma}_j(t')\rangle$ ($i=1,2,j=3,4$) vanish. By substituting \eqref{operator_new_picture} into the generalized master equation (\ref{gme}), utilizing the bath correlations to simplify the integrals, and finally transferring it back to the Schr\"{o}dinger picture, we derive
\begin{equation}\label{me_2B}
\frac{d\hat{\rho}}{dt}=i[\hat{\rho},\hat{\mathcal{H}}_{\mathrm{AFM}}] + \mathcal{L}^{(0)}[\hat{\rho}],
\end{equation}
where the Liouville operator $\mathcal{L}^{(0)}[\hat{\rho}]$ reads
\begin{widetext}
\begin{equation}\label{liouville_2B}
\begin{aligned}
\mathcal{L}^{(0)}[\hat{\rho}]=(n_{\mathrm{th}}+1)\kappa \left[\eta_{\omega\omega} \mathcal{L}_{aa}[\hat{\rho}]+\eta_{\omega g}(\mathcal{L}_{ab^\dagger}[\hat{\rho}] + \mathcal{L}_{b^\dagger a}[\hat{\rho}])+\eta_{gg}\mathcal{L}_{b^\dagger b^\dagger} [\hat{\rho}] \right ]\\
+ n_{\mathrm{th}}\kappa\left[\eta_{\omega\omega} \mathcal{L}_{a^\dagger a^\dagger}[\hat{\rho}]+\eta_{\omega g}(\mathcal{L}_{ba^\dagger}[\hat{\rho}] + \mathcal{L}_{a^\dagger b})[\hat{\rho}]+\eta_{gg} \mathcal{L}_{bb} [\hat{\rho}]\right ]\\
+(n_{\mathrm{th}}+1)\kappa \left[\eta_{\omega\omega} \mathcal{L}_{bb}[\hat{\rho}]+\eta_{\omega g} (\mathcal{L}_{ba^\dagger}[\hat{\rho}] + \mathcal{L}_{a^\dagger b}[\hat{\rho}])+\eta_{gg} \mathcal{L}_{a^\dagger a^\dagger} [\hat{\rho}] \right ]\\
+ n_{\mathrm{th}}\kappa \left[\eta_{\omega\omega} \mathcal{L}_{b^\dagger b^\dagger}[\hat{\rho}]+\eta_{\omega g}(\mathcal{L}_{ab^\dagger}[\hat{\rho}] + \mathcal{L}_{b^\dagger a})[\hat{\rho}]+\eta_{gg}\mathcal{L}_{aa} [\hat{\rho}]\right ],\\
\end{aligned}
\end{equation}
\end{widetext}
where the relaxation rate has the same definition as that of Eq. \eqref{singlemode_me}. Here the relaxation coefficients are slightly different from the single-mode case, i.e.,
\begin{equation}
\eta_{\omega \omega} = \frac{(\omega_r+\omega_a)^2}{4\omega_r^2},
\eta_{\omega g} = \frac{ g_{ab}(\omega_r+\omega_a)}{4\omega_r^2},
\eta_{gg} = \frac{ g_{ab}^2}{4\omega_r^2}.
\end{equation}
The dependence of these relaxation coefficients on the coupling strength of two magnons ($g_{ab}$) is shown in Fig. \ref{fig2}(a). For a crystalline AFM, the exchange field usually dominates the anisotropy ($H_{\mathrm{ex}} \gg H_{\mathrm{an}}$), i.e., $g_{ab}/\omega_a$ is close to 1, then the three relaxation coefficients are comparable in strength and all of them should be taken into account. For a synthetic AFM and van der Waals bilayered magnet, the two types of magnons excited on each layer are coupled through the Ruderman-Kittel-Kasuya-Yosida (RKKY) or van der Waals interaction, which is tunable over a large scale by layer distance and electric field \cite{RembertNP2018,ZhangNM2020}. Then the collective relaxation channels $\eta_{\omega g}$ and $\eta_{g g}$ may not play a significant role in the regime $H_{\mathrm{ex}} \ll H_{\mathrm{an}}$ ($g_{ab}/\omega_a \sim 0$). The intuitive picture to understand the local and collective dissipation channels is plotted in Fig. \ref{fig2}(b). Instead of single-mode squeezing as shown in Fig. \ref{fig1}(b), now the two-mode squeezing will launch the collective dissipation channels after simultaneously exciting two magnons.

Equipped with the generalized master equation \eqref{me_2B}, we study the collective dissipation channel's influence on the steady state of the system. We first simulate the steady magnon occupation and entanglement of sublattice magnons as a function of the magnon-magnon coupling and show the results in Fig. \ref{fig2}(c). Here the entanglement of sublattice magnons $E_n$ is quantified as the logarithmic negativity of the joint density matrix \cite{Yuanreview2022}. In the ground state, $E_N=r$ (black dashed line in Fig. \ref{fig2}(c)), where $r$ is nothing but the squeezing parameter of the two-mode squeezed state. At zero temperature, the system will gradually evolve toward this ground state, when all the dissipation channels are included (red solid line). This is physically expected. If only the local relaxation channels are considered, however, the system will end up in a different steady states with lower entanglement (red dashed lines), similar to the relaxation of single-mode squeezed state as discussed in Sec. \ref{sec_single_mode}. The collective dissipation channel becomes more pronounced for stronger exchange coupling, which is typically valid for a crystalline antiferromagnet. Through this comparison, it is clear that the inclusion of the collective dissipation channel is indispensable to recover the physical steady state of the system.

When thermal fluctuations are taken into account, the entanglement of sublattice magnons decreases and is subject to a sudden death at a critical temperature (around 8 K for $\omega_r =1$ THz), as shown in Fig. \ref{fig2}(d). Here, the collective dissipation channel makes the magnon-magnon entanglement survive at a even higher temperature (red solid line) and thus increases the thermal stability of the entanglement. This also resembles the behavior of single-mode squeezed state.

\subsection{One common bath}\label{sec_afm_onebath}

Now we proceed to consider the case in which the two types of magnons share one bath.
One common bath implies that the bath operators of the second bath are the same as the first one in Eq. \eqref{gme}, i.e., $\tilde{\Gamma}_3(t)=\tilde{\Gamma}_1(t), \tilde{\Gamma}_4(t)=\tilde{\Gamma}_2(t)$ and thus the cross correlations terms $\langle \tilde{\Gamma}_1(t)\tilde{\Gamma}_4(t') \rangle$ and $\langle \tilde{\Gamma}_2(t)\tilde{\Gamma}_3(t') \rangle$ will not vanish any longer and they will add additional dissipation channels to the system. Following similar procedures as Sec. \ref{sec_afm_twobath}, we obtain
\begin{equation}\label{me_1B}
\frac{d\hat{\rho}}{dt}=i[\hat{\rho},\hat{\mathcal{H}}_{\mathrm{AFM}}]+\mathcal{L}^{(0)} [\hat{\rho}] + \mathcal{L}^{(1)} [\hat{\rho}],
\end{equation}
where the Liouville operator $\mathcal{L}^{(0)}[\hat{\rho}]$ takes the same form as Eq. \eqref{liouville_2B}, while $\mathcal{L}^{(1)}[\hat{\rho}]$ is purely caused by the introduction of the common bath and it reads
\begin{widetext}
\begin{equation}\label{liouville_1B}
\begin{aligned}
\mathcal{L}^{(1)} [\hat{\rho}] &=\kappa (n_{\mathrm{th}}+1) \left[\eta_{\omega\omega} (\mathcal{L}_{ab}+\mathcal{L}_{ba})[\hat{\rho}]+\eta_{\omega g}(\mathcal{L}_{aa^\dagger}+ \mathcal{L}_{a^\dagger a} + \mathcal{L}_{b b^\dagger} + \mathcal{L}_{b^\dagger b})[\hat{\rho}]+\eta_{gg}(\mathcal{L}_{b^\dagger a^\dagger}+\mathcal{L}_{a^\dagger b^\dagger}) [\hat{\rho}] \right ]\\
&+\kappa n_{\mathrm{th}} \left[\eta_{\omega\omega} (\mathcal{L}_{b^\dagger a^\dagger} + \mathcal{L}_{a^\dagger b^\dagger})[\hat{\rho}]+\eta_{\omega g}(\mathcal{L}_{aa^\dagger}+ \mathcal{L}_{a^\dagger a} + \mathcal{L}_{b b^\dagger} + \mathcal{L}_{b^\dagger b})[\hat{\rho}]+\eta_{gg}(\mathcal{L}_{ab}+\mathcal{L}_{ba}) [\hat{\rho}] \right ].\\
\end{aligned}
\end{equation}
\end{widetext}
When the coupling between the two magnons is zero, i.e., $g_{ab}=0$, the master equation \eqref{me_1B} is reduced to the form
\begin{widetext}
\begin{equation}\label{megab0}
\frac{d\hat{\rho}}{dt}=i[\hat{\rho},\hat{\mathcal{H}}_{\mathrm{AFM}}] + \kappa (n_{\mathrm{th}}+1)(\mathcal{L}_{aa}+\mathcal{L}_{bb}+\mathcal{L}_{ab}+\mathcal{L}_{ba})[\hat{\rho}]+\kappa n_{\mathrm{th}}(\mathcal{L}_{a^\dagger a^\dagger}+\mathcal{L}_{b^\dagger b^\dagger}+\mathcal{L}_{a^\dagger b^\dagger}+\mathcal{L}_{b^\dagger a^\dagger})[\hat{\rho}].
\end{equation}
\end{widetext}
This master equation describes the evolution of any non-interacting boson system, and is not limited to magnonic system. It is also straightforward to extend it to the multi-mode case, by noticing the symmetry of boson modes distribution in the Liouville operator.
As a comparison, the evolution of two qubits coupled to a common bath has been more widely discussed, because it supports decoherence free evolution \cite{ZanardiPRL1996,DuanPRL1997,LidarPRL1998,KwiatScience2000} and is useful for quantum computing without error correction. We will show that this is also true for both non-interacting and interacting magnons.

Let us first analyze how a physical observable evolves under the master equation \eqref{me_1B}. Note that the master equation \eqref{me_1B} is linear in the density matrix and it can be formally written as,
$d \hat{\rho}/dt = \mathcal{L}\hat{\rho}$, which has a formal solution $\hat{\rho}(t) = e^{\mathcal{L} t} \hat{\rho}(0)$. By defining an adjoint Liouville operator as $tr(\hat{A}\mathcal{L}(\hat{B}))= tr(\mathcal{L}^\dagger(\hat{A})\hat{B})$, we derive the equation of motion for the ensemble average of any operator $\hat{A}$ as $d\langle \hat{A} \rangle/dt=\langle \mathcal{L}^\dagger(\hat{A}) \rangle$, or, equivalently
\begin{widetext}\label{meAgab0}
\begin{equation}
\frac{d \langle \hat{A} \rangle}{dt}=-i\langle [\hat{A},\hat{\mathcal{H}}_{\mathrm{AFM}}] \rangle + \mathcal{D}^{(0)} [\hat{\rho}] +\mathcal{D}^{(1)} [\hat{\rho}],
\end{equation}
\end{widetext}
where $\mathcal{D}^{(0)} [\hat{\rho}]$ and $\mathcal{D}^{(1)} [\hat{\rho}]$ is converted from the Lindblad operators in \eqref{liouville_2B} and \eqref{liouville_1B} following the rule,
$\mathcal{D}_{AB}(\hat{O})= \langle 2\hat{B}^\dagger \hat{O}\hat{A} - \hat{A}^\dagger \hat{B} \hat{O} - \hat{O}\hat{A}^\dagger \hat{B} \rangle$.
%\begin{widetext}\label{meAgab0}
%\begin{equation}
%\frac{d \langle \hat{A} \rangle}{dt}=-i\langle [\hat{A},\hat{\mathcal{H}}_{\mathrm{AFM}}] \rangle + \kappa (n_{\mathrm{th}}+1)(\mathcal{D}_{aa}+\mathcal{D}_{bb}+\mathcal{L}_{ab}+\mathcal{D}_{ba})[\hat{A}]+n_{\mathrm{th}}(\mathcal{D}_{a^\dagger a^\dagger}+\mathcal{D}_{b^\dagger b^\dagger}+\mathcal{D}_{a^\dagger b^\dagger}+\mathcal{D}_{b^\dagger a^\dagger})[\hat{A}]
%\end{equation}
%\end{widetext}
Based on this equation, we derive the evolution of magnon operators $\hat{a}$ and $\hat{b}$ as
\begin{widetext}
\begin{subequations}\label{eqab1Bath}
\begin{align}
\frac{d\langle \hat{a} \rangle}{dt} &= -i\left [\omega_a-i(\eta_{\omega\omega} -\eta_{gg}) \kappa \right ] \langle \hat{a} \rangle - ig_{ab} \langle \hat{b}^\dagger \rangle - (\eta_{\omega\omega}-\eta_{gg})\kappa \langle \hat{b} \rangle,\\
\frac{d\langle \hat{b} \rangle}{dt} &= -i\left [\omega_b-i(\eta_{\omega\omega} - \eta_{\omega\omega}) \kappa \right ] \langle \hat{b} \rangle- ig_{ab} \langle \hat{a}^\dagger \rangle - (\eta_{\omega\omega} -\eta_{\omega\omega})\kappa \langle \hat{a} \rangle.
\end{align}
\end{subequations}
\end{widetext}

There are two more equations on the evolution of $\langle \hat{a}^\dagger \rangle$ and $\langle \hat{b}^\dagger \rangle$, which are obtained by taking the Hermitian conjugate on both sides of Eqs. \eqref{eqab1Bath}. By solving this set of equations, we can derive the coherence of magnon mode as
%\begin{widetext}
%\begin{equation}
%|\langle a(t) \rangle|=\frac{1}{2 \omega_r}\left| \begin{array}{c}
%                                                 ig_{ab}\left [(a^\dagger(0)+b^\dagger (0)) e^{-(\kappa+\kappa_c) t} - (a^\dagger(0)-b^\dagger(0))e^{-(\kappa-\kappa_c) t})\sin (\omega_{r}t) \right ] \\
%                                                 - \left [ (a(0)+b(0))e^{-(\kappa+\kappa_c) t} + (a(0)-b(0))e^{-(\kappa-\kappa_c) t}\right ] \left [ \omega_r \cos (\omega_r t) - i\sqrt{\omega_r^2 + g_{ab}^2}\sin (\omega_r t) \right ]
%                                               \end{array}
% \right|
%\end{equation}
%\end{widetext}
%where $\eta=\kappa_{\omega\omega}^{(2)}-\kappa_{gg}^{(2)}$, $\eta_c=\kappa_{\omega\omega}^{(1)}-\kappa_{\omega\omega}^{(1)}$.
\begin{widetext}
\begin{equation}\label{at_evolution}
|\langle \hat{a}(t) \rangle|=\frac{1}{2 \omega_r}\left| \begin{array}{c}
                                                 ig_{ab}\left [\langle \hat{a}^\dagger(0)+\hat{b}^\dagger (0) \rangle e^{-2(\eta_{\omega \omega}-\eta_{gg})\kappa t} - \langle \hat{a}^\dagger(0)-\hat{b}^\dagger(0)\rangle )\sin (\omega_{r}t) \right ] \\
                                                 - \left [ \langle \hat{a}(0)+\hat{b}(0)\rangle e^{-2(\eta_{\omega \omega}-\eta_{gg})\kappa t} + \langle \hat{a}(0)-\hat{b}(0) \rangle \right ] \left [ \omega_r \cos (\omega_r t) - i\sqrt{\omega_r^2 + g_{ab}^2}\sin (\omega_r t) \right ]
                                               \end{array} \right|.
\end{equation}
\end{widetext}

Since $\eta_{\omega \omega}>\eta_{gg}$ as shown in Fig. \ref{fig2}(a),  $|\langle a(t) \rangle|$ will always decay on the time scale of $\tau_{qm}=1/[(\eta_{\omega \omega}-\eta_{gg})\kappa]$. After a sufficiently long time, we have
\begin{widetext}
\begin{equation}
|\langle \hat{a}(t) \rangle|=\frac{1}{2 \omega_r}\left|
ig_{ab}\left [\langle\hat{a}^\dagger(0)-\hat{b}^\dagger(0)\rangle\sin (\omega_{r}t) \right ] \\
                                                 + \langle \hat{a}(0)-\hat{b}(0) \rangle \left [ \omega_r \cos (\omega_r t) - i\sqrt{\omega_r^2 + g_{ab}^2}\sin (\omega_r t) \right ] \right|.
\end{equation}
\end{widetext}
In the non-interacting case, $g_{ab}=H_{\mathrm{ex}}=0$, $|\langle \hat{a}(t) \rangle | = |\hat{a}(0)-\hat{b}(0)|$. This result implies that, given an initial coherent state reading
$| \psi_0 \rangle =|\beta \rangle \otimes |-\beta \rangle$ ($\beta >0 $ for example), the magnons will stay in this state without any decoherence ($|\langle a(t) \rangle| = \beta$). Hence, this is a decoherence free evolution.
In the limit with $g_{ab}=H_\mathrm{ex} \gg H_\mathrm{an}$, we obtain $\omega_r \ll g_{ab}$, $|\langle a(t) \rangle| =\beta |\cos(\omega_r t)|$, implying that the coherence will keep oscillating with the period of $2\pi/\omega_r$. In the intermediate regime, we have
\begin{equation}
|\langle \hat{a}(t) \rangle|=\beta/\sqrt{2} \sqrt{(1+\zeta^2) +(1-\zeta^2) \cos 2 \omega_r t},
\end{equation}
where $\zeta = \sqrt{1+(g_{ab}/\omega_r)^2} - g_{ab}/\omega_r$. Now $|\langle a(t) \rangle|$ will oscillate around the expectation value $2\sqrt{2}\beta K(1-\zeta^2)$ with $K(x)$ being the complete elliptic integral of the first kind.

\begin{figure}
  \centering
  \includegraphics[width=0.45\textwidth]{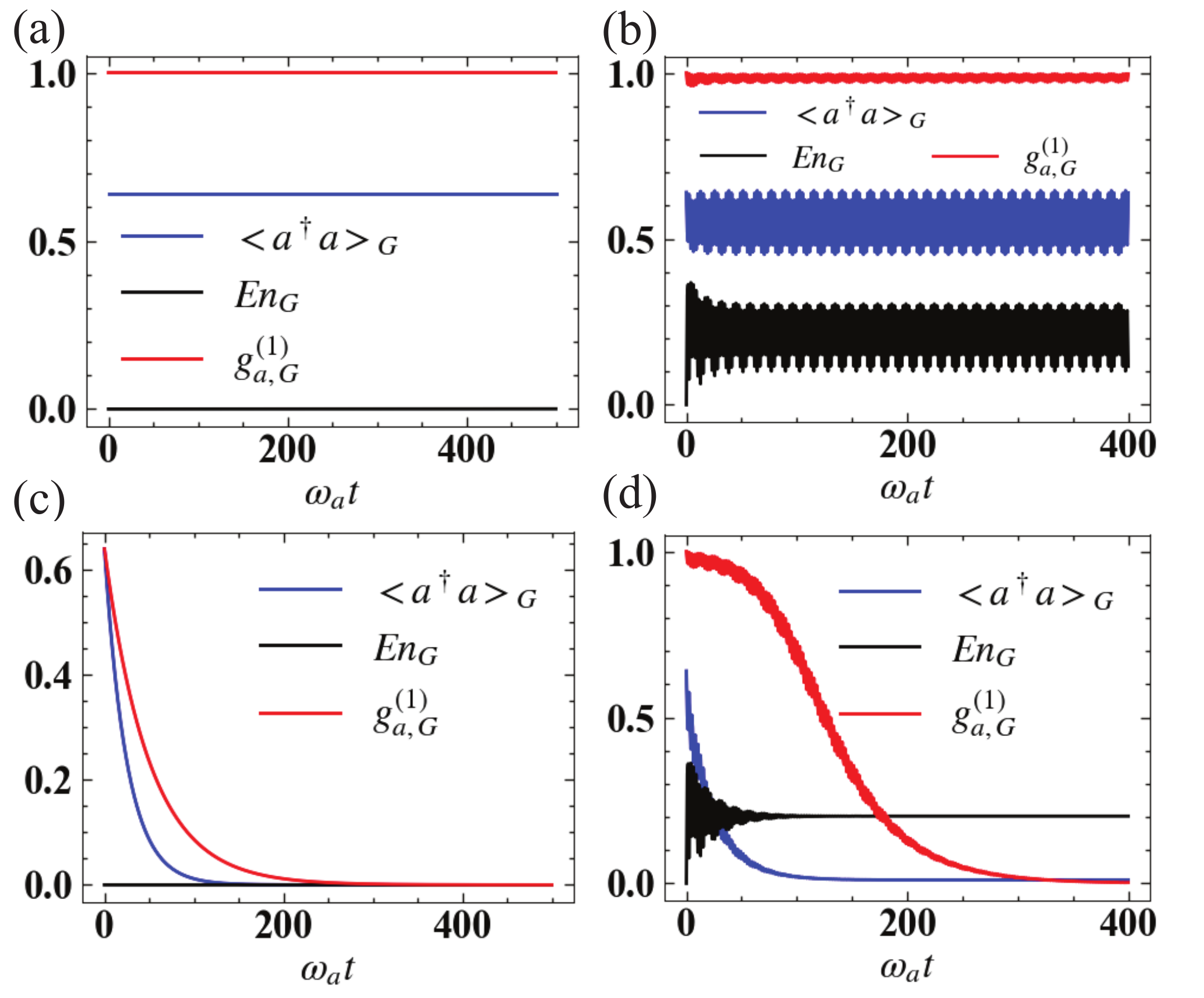}\\
  \caption{Magnon occupation $\langle a^\dagger a\rangle_G$, magnon-magnon entanglement $E_{nG}$ and first-order coherence of $a-$type magnon $g^{(1)}_{a,G}$ as a function of time at zero temperature. $|\psi_0\rangle = |\beta \rangle \otimes |-\beta \rangle, \beta =0.8$. (a) Decoherence free evolution when the sublattice magnons are decoupled but share one common bath. $ \kappa_1=0.04, \kappa_2=0, g_{ab}=0$. (b) Oscillatory behavior of the magnon coherence when the sublattice magnons are strongly coupled and share one common bath. $\kappa_1=0.04, \kappa_2=0,g_{ab}=0.2\omega_a$. (c) Decoherence of the magnons when the sublattice magnons are decoupled but share a hybrid bath. $\kappa_2=0.04, \kappa_1=0.004,g_{ab}=0$. (d) Decoherence of the magnons when the sublattice magnons are coupled and share a hybrid bath. Now a steady entanglement of sublattice magnons survives resulted from the squeezed ground state of the system. $\kappa_2=0.04, \kappa_1=0.004,g_{ab}=0.2\omega_a$ \cite{noteg1}.}\label{fig3}
\end{figure}

To verify these analytical predictions, we simulate the dynamics of the system by numerically solving the master equation \eqref{me_1B} and show the results in Fig. \ref{fig3}(a) and (b). Here we numerically evaluate both the magnon occupation $\langle \hat{a}^\dagger \hat{a}\rangle$ and first-order coherence function of magnons as, $g^{(1)}_a = |\hat{a}^\dagger (t) \hat{a}(0)|$. Clearly, in the absence of magnon coupling ($g_{ab}=0$), there is a decoherence free evolution, while the decoherence oscillates in time when the coupling of two magnon modes is included ($g_{ab} \neq 0$).

%\subsection{Hybrid bath}
%
%Now we can change the intersublattice coupling in Eq. (\ref{eqab1Bath}) becomes $\kappa_c$, and time dependence of coherence reads,
%
%\begin{widetext}
%\begin{subequations}
%\begin{align}
%&|\langle a^\dagger(t)a(0)|=\frac{1}{2}\left|(a^\dagger(0)a(0)+b^\dagger(0)a(0))e^{-(\kappa + \kappa_c) t}+(a^\dagger(0)a(0)-b^\dagger(0)a(0))e^{-(\kappa - \kappa_c)t} \right|,\\
%&\langle a^\dagger(t)a(t)\rangle = \frac{1}{4}\left[ (N_0-2n_{th})e^{-2(\kappa + \kappa_c)t} +2(a^\dagger(0)a(0)-b^\dagger(0)b(0))e^{-2\kappa t}+(I_0+2n_{th})e^{-2(\kappa - \kappa_c)t} \right ].
%\end{align}
%\end{subequations}
%\end{widetext}
%where $N_0 = \langle a^\dagger(0)a(0)+b^\dagger(0)b(0)+a^\dagger(0)b(0)+b^\dagger(0)a(0) \rangle$ and $I_0= \langle a^\dagger(0)a(0)+b^\dagger(0)b(0)-a^\dagger(0)b(0)-b^\dagger(0)a(0)\rangle$. We immedately see that the magnon ocupation and unnormalized coherence will decay to zero with the time scale $1/(\kappa-\kappa_c)$. The dissipation channel $\kappa_c$ via the one common bath enhances the magnon lifetime. This is consistent with the classical relaxation of spins, by introducing the intersublattice damping. To make this point clear, let us reacall the classical dynamic equations of the system \cite{Yuan20172},
%

\subsection{Classical-quantum correspondence}
We have shown the relaxation behavior of antiferromagnetic magnons within the framework of quantum master equation approach. This formalism allows us to study the quantum properties of magnons, for exmaple, magnon squeezing and entanglement. In modern spintronics, a widely used approach is to study the magnon excitation and propagation based on the classical Landau-Lifshitz-Gilbert (LLG) equation \cite{Landau1935,Gilbert2004}. In this classical formalism, the relaxation of magnons is modelled by a phenomenological parameter called Gilbert damping. In this section, we compare the relaxation rate in the quantum equation of motion with that in the classical equation of motion, and show that the models of two separate baths and one common bath each capture part of the relaxation process, while a hybrid bath model seems more proper to fully describe the magnon relaxation. Even with this correspondence, we emphasize that the quantum formalism is essential to study the quantum properties, such as entanglement, of magnons. These cannot be addressed using the classical formalism.

\textit{Classical equation of motion.---} The dissipation channel via one common bath enhances the magnon lifetime. This is consistent with the classical relaxation of spins, by introducing the intersublattice damping. To make this point clear, let us recall the LLG equations of a two-sublattice AFM \cite{YuanEPL2019}

\begin{equation}
\frac{\partial \mathbf{S}_j}{\partial t}=-\mathbf{S}_j\times \mathbf{H}_j
+ \mathbf{S}_j \times \left [\alpha\frac{\partial \mathbf{S}_j}{\partial t}
+ \frac{\alpha_c}{2} \left ( \frac{\partial \mathbf{S}_{j-1}}{\partial t}
+ \frac{\partial \mathbf{S}_{j+1}}{\partial t} \right ) \right ],
\label{2llg}
\end{equation}
where $\mathbf{H}_j=-\delta \mathcal{H}_{\mathrm{AFM}}/\delta \mathbf{S}_j$ is the effective field acting on the spin of $j-$th site. Here $\mathbf{S}_j$ is a classical vector with magnitude $S$, instead of the quantum operator in the quantum model.
The first and second terms on the right-hand-side of Eq. \eqref{2llg} describe the spin precessing and relaxation toward the equilibrium axis, respectively. The coefficients $\alpha$ and $\alpha_c$ are two phenomenological parameters that characterize the strength of dissipation, and they are usually termed as Gilbert damping and intersublattice damping \cite{YuanEPL2019,KamraPRB2018}. Recently, 
Equation \eqref{2llg} was reproduced using a Caldeira-Leggett approach when the two subsystems have both shared and separate baths \cite{DaanarXiv2022}.

In the classical picture, the lowest-energy magnon excitation can be viewed as the uniform precession of spins around the N\'{e}el-type ground state, i.e., $\mathbf{S}_{2j}=Se_z + \delta \mathbf{S}_a(t),~\mathbf{S}_{2j+1}=-Se_z+
\delta \mathbf{S}_b(t)$. By substituting the trial solutions into Eq. (\ref{2llg}) and keeping
only the linear terms in the small perturbations $\delta \mathbf{S}_{a,b}$, we obtain
\begin{equation}
i\frac{\partial }{\partial t} \left (\begin{array}{c}
                                \delta S_a^+ \\
                                \delta S_b^+
                              \end{array} \right )
                              =\mathcal{H}_{\mathrm{eff}}  \left (\begin{array}{c}
                                \delta S_a^+ \\
                                \delta S_b^+
                              \end{array} \right ),
\label{ptHam}
\end{equation}
where $\delta S_{a,b}^+=\delta S_{a,b}^x+i\delta S_{a,b}^y$ are classical fields, corresponding to the spin rising operators in the quantum framework, and $\mathcal{H}_{\mathrm{eff}}$ is the effective Hamiltonian of magnon excitation that takes the form
\begin{widetext}
\begin{equation}\label{Ham_cl_afm}
\mathcal{H}_{\mathrm{eff}} = \left ( \begin{array}{cc}
         -\omega_a+i(\alpha_c g_{ab} - \alpha \omega_a) & -g_{ab}+i(\alpha_c \omega_a - \alpha g_{ab}) \\
         g_{ab}+i(\alpha_c \omega_a - \alpha g_{ab}) & \omega_a+i(\alpha_c g_{ab} - \alpha \omega_a)
       \end{array} \right ).
\end{equation}
\end{widetext}
Here we have assumed $\alpha,\alpha_c \ll 1$ to safely neglect the higher-order terms of damping. The Hamiltonian \eqref{Ham_cl_afm} is non-Hermitian for the existence of damping in the dynamic equations, but $\mathcal{PT}$ antisymmetric ($\mathcal{APT}$). To prove this statement,
 we introduce the parity and time reversal operators as follows
 \begin{equation}
   \hat{\mathcal{P}}=\hat{\sigma}_x,~ \hat{\mathcal{T}}=\hat{\mathcal{I}} \hat{\mathcal{K}},
 \end{equation}
where $\hat{\sigma}_x$ is Pauli matrix which exchanges the position of two types of magnons ($a\leftrightarrow b$), $\hat{\mathcal{I}}$
is identify operator and $\hat{\mathcal{K}}$ is complex conjugate. One can readily prove
$\{\mathcal{\hat{P}\hat{T}},\mathcal{H}_{\mathrm{eff}}\}=0$ such that the system is $\mathcal{PT}$ antisymmetric.
Now two phases can be classified according to the eigenvalues ($\omega_\pm$) of $\mathcal{H}_{\mathrm{eff}}$. If both eigenvalues are purely imaginary, the $\mathcal{APT}$ symmetry is unbroken, which is denoted as
$\mathcal{APT}$ exact phase. Otherwise, the $\mathcal{APT}$ symmetry is broken.

Specifically, the eigenvalues of the system can be obtained by solving the secular equation $\det(\omega \mathcal{I}-\mathcal{H}_{\mathrm{eff}} )=0$ as
\begin{equation}
\omega_\pm = -i(\alpha \omega_a - \alpha_c g_{ab} ) \pm \omega_r.
\end{equation}
Now the phase boundary separating the $\mathcal{APT}$ exact and and $\mathcal{APT}$ broken phase is $\alpha_c= \alpha \omega_a/g_{ab}$.
In the $\mathcal{APT}$ exact phase, $\mathrm{Im}(\omega_{\pm})>0$, then the linearized fluctuations will grow exponentially and diverge, which indicates that
the N\'{e}el-type ground state is not stable any longer. Here we will always focus our discussions on the $\mathcal{APT}$ broken phase. In this regime, the magnon lifetime
is $\tau_{cl} = 1/(\alpha \omega_a-\alpha_c g_{ab})$, corresponding to the magnon relaxation time $\tau_{qm} = 1/[(\eta_{\omega \omega} - \eta_{gg})\gamma]$ in the quantum formalism.

\textit{Classical vs quantum.---} Now let us go a step further to connect the classical and quantum relaxation process. By comparing the classical equation of motion (\ref{ptHam}) with the quantum equation (\ref{eqab1Bath}) obtained from the one bath model, we find the correspondence
\begin{equation}
\alpha \omega_a \leftrightarrow \eta_{\omega \omega} \kappa,~\alpha_c g_{ab} \leftrightarrow \eta_{gg} \kappa.
\end{equation}
Note that $\eta_{\omega \omega} > \eta_{gg}$ as shown in Fig. \ref{fig2}(a), implying $\alpha \omega_a>\alpha_cg_{ab}$, hence the stability of the N\'{e}el-type ground state is always guaranteed. On the other hand, if we compare the classical equation of motion (\ref{ptHam}) with the quantum equation obtained from the two-bath model, we find the correspondence
\begin{equation}
\alpha \omega_a \leftrightarrow \eta_{\omega \omega} \kappa,~\alpha_c g_{ab} \leftrightarrow 0.
\end{equation}

Now we are in a position to discuss which quantum model is more proper to describe the magnon relaxation in an AFM.
If the intersublattice damping of a material is zero, within the master equation approach, it seems that we can readily choose the two bath model, which means that the sublattice magnons respectively dissipate angular momentum into their own bath and the spin information emitted by one sublattice magnon will quickly lose in the bath and cannot influence the precession of the other magnon. This may be true for the recent 2D bilayered van der Waals magnets,
where the observed magnon linewidth does not support the prediction of intersublattice damping \cite{ZhangNM2020}, and synthetic AFMs. On the other hand, if the intersublattice damping of a material is not zero, for example in the metallic AFMs, $\mathrm{XMn}$ ($\mathrm{X=Pt,Pd,Ir,Fe}$) as confirmed by first-principles calculations \cite{LiuPRM2017,MahfouziPRB2018}, the one bath model must be taken into account. One immediate question is whether the one bath model is sufficient to describe such an AFM.

The answer is no, which can be justified by \textit{reductio ad absurdum}. If the two sublattice magnons perfectly share one bath, then the coherent precession of spins will persist forever as discussed in the last section. This may be the case if one can artificially engineer the bath perfectly, but is apparently not true in a natural AFM according to the second law of thermodynamics. Combing all these discussions, it seems more reasonable to use a hybrid bath model to completely describe the magnon dissipation. Here the spins on the two sublattices each have an independent channel to relax, for example, by coupling to the phonons and impurities, while they also share a common bath, for example, the conduction electrons could bridge the dissipation of the two sublattices, as justified by the first-principle calculations \cite{LiuPRM2017} and analytical treatment of magnon-electron scattering \cite{SimensenPRB2020}. Taking all these into account, the master equation reads
\begin{equation}
\frac{d\hat{\rho}}{dt} = i[\hat{\rho},\hat{\mathcal{H}}_{\mathrm{AFM}}] + \left [\mathcal{L}^{(0,2B)} + \mathcal{L}^{(0,1B)}  + \mathcal{L}^{(1,1B)} \right ][\hat{\rho}],
\end{equation}
where $\mathcal{L}^{(0,2B)}$ and $\mathcal{L}^{(0,1B)}$ have identical form as Eq. (\ref{liouville_2B}), while $\mathcal{L}^{(1,1B)}$ has the form Eq. (\ref{liouville_1B}).

Following the same procedures to derive Eq. \eqref{eqab1Bath}, we find the correspondence principle,
\begin{equation}
\alpha \omega_a \leftrightarrow \eta_{\omega \omega} (\kappa_2+\kappa_1),~\alpha_c g_{ab} \leftrightarrow \eta_{gg} \kappa_1,
\end{equation}
where $\kappa_2$ and $\kappa_1$ are respectively the coupling strength of the spins to the two-separate bath and one common bath.
Now the perfect decoherence-free evolution predicted in the model of one common bath vanishes \cite{DaanarXiv2022} and $|\langle a(t) \rangle|$ will decay to zero with a time scale of $\tau^{-1}=(\eta_{\omega \omega}-\eta_{gg}) \kappa_{2}$. The strength of $\kappa_2$ directly determines the decoherence time of the system, which corresponds to the intrinsic damping stated in Ref. \cite{YuanEPL2019}.

Typical evolutions of particle number and coherence of the hybrid model are shown in Fig. \ref{fig3}(c) and (d). Now the coherence of an initial product state gradually decreases to zero, regardless of the strength of coupling between sublattice magnons ($g_{ab}$).  The particle number evolves to zero (a finite value) without (with) the sublattice coupling. The finite sublattice coupling also generates a steady entanglement between the magnons, as expected. This is different from the oscillating behavior of entanglement shown in Fig. \ref{fig3}(b).

\section{Pure dephasing of magnons}\label{sec_dephase}
Besides the relaxation of magnons as discussed in Sec. \ref{sec_single_mode} and \ref{sec_two_mode}, pure dephasing of magnons without energy relaxation is another important channel that can cause decoherence of the magnons. In our recent work \cite{YuanArxiv2022}, we have explicitly demonstrated that both four-magnon scattering and magnon-phonon scattering contribute to the pure dephasing of magnons, and the dephasing rate depends on the exchange coefficient, magnetoelastic coupling strength, external field and temperature. In this section, we provide a detailed derivation of the master equation accounting for pure dephasing, following a methodology similar to our considerations above.

\subsection{Four-magnon scattering}
We consider an isotropic Heisenberg magnet subject to an external magnetic field, which is described by Eq. \eqref{Ham_FM} with $K_x=K_z=0$. Employing the HP transformation of the spin operators and retaining up to the fourth-order terms of magnon-magnon interaction, we derive the effective Hamiltonian of magnon excitations as \cite{YuanArxiv2022},
\begin{equation} \label{heisenberg_ham}
\hat{\mathcal{H}}_\mathrm{FM}=\sum_\mathbf{k} \omega_\mathbf{k} \hat{a}_\mathbf{k}^\dagger \hat{a}_\mathbf{k} + \sum_{\mathbf{k},\mathbf{k}',\mathbf{q}} C(\mathbf{k},\mathbf{k}',\mathbf{q}) \hat{a}_\mathbf{k+q}^\dagger \hat{a}_\mathbf{k'-q}^\dagger \hat{a}_\mathbf{k'} \hat{a}_\mathbf{k},
\end{equation}
where $\omega_\mathbf{k}=ZJSd^2\mathbf{k}^2+H$ is the magnon dispersion relation with $d$ the lattice constant, $C(\mathbf{k},\mathbf{k}',\mathbf{q})$ is the interaction strength of magnons that depends on the exchange coefficients and structure factor of the system. Different from the consideration in Sec. \ref{sec_single_mode} and \ref{sec_two_mode}, now both uniform ($\mathbf{k}=0$) and nonuniform ($\mathbf{k} \neq 0$) precession modes as well as their interaction are taken into account. This is essential to discuss the dephasing of magnons caused by exchange interaction. As we go further, we simplify the triple sum on the right-hand-side of Eq. \eqref{heisenberg_ham} by focusing on the dynamics of a particular magnon mode with wavevector $\mathbf{k}_0$ (for example, the ferromagnetic resonance mode). Then we reduce the four-magnon interaction in \eqref{heisenberg_ham} and rewrite the Hamiltonian as
\begin{equation} \label{exchange_ham}
\hat{\mathcal{H}}_\mathrm{FM}=\omega_r \hat{a}^\dagger \hat{a} +\sum_{\mathbf{k} \neq \mathbf{k}_0} \omega_\mathbf{k} \hat{a}_\mathbf{k}^\dagger \hat{a}_\mathbf{k} + \hat{a}^\dagger \hat{a} \sum_{\mathbf{k} \neq \mathbf{k}_0} g(\mathbf{k})\hat{a}_\mathbf{k}^\dagger \hat{a}_\mathbf{k},
\end{equation}
where we have defined $\hat{a} \equiv \hat{a}_{\mathbf{k}_0},\omega_r= \omega_{\mathbf{k}_0}, g(\mathbf{k}) = C(\mathbf{k}_0,\mathbf{k},\mathbf{q}=0)$ to make the notation simple and consistent with our notations in discussing magnon relaxation.

Following the discussion in Ref. \cite{YuanArxiv2022}, we shall treat all the magnon modes ($\mathbf{k}\neq \mathbf{k}_0$) as bath degrees of freedom and trace them out to derive the dynamics of the magnon mode $\mathbf{k}_0$. This treatment readily allows us to apply our generalized master equation approach presented in Sec. \ref{general_me} to study this problem. Recalling the generalized master equation \eqref{gme}, we immediately have the correspondence
$\tilde{s}=\hat{a}^\dagger \hat{a}$ and $\tilde{\Gamma} = \sum_\mathbf{k} g(\mathbf{k})\hat{a}_\mathbf{k}^\dagger \hat{a}_\mathbf{k}$. However, we cannot apply this correspondence directly, because the derivation of Eq. \eqref{gme} requires the mean average of the bath operator being equal to zero, i.e., $\langle \tilde{\Gamma} \rangle =0$, which is not true here. To fulfill this requirement, we reformulate the Hamiltonian \eqref{exchange_ham} as

\begin{equation} \label{exchange_ham2}
\hat{\mathcal{H}}_\mathrm{FM}=\omega_r^{'} \hat{a}^\dagger \hat{a} +\sum_{\mathbf{k} \neq \mathbf{k}_0} \omega_\mathbf{k} \hat{a}_\mathbf{k}^\dagger \hat{a}_\mathbf{k} + \hat{a}^\dagger \hat{a} \sum_{\mathbf{k} \neq \mathbf{k}_0} g(\mathbf{k}) ( \hat{a}_\mathbf{k}^\dagger \hat{a}_\mathbf{k} -\langle \hat{a}_\mathbf{k}^\dagger \hat{a}_\mathbf{k} \rangle ),
\end{equation}
where $\omega_r^{'}= \omega_r+\sum_k  g(\mathbf{k}) \langle \hat{a}_\mathbf{k}^\dagger \hat{a}_\mathbf{k} \rangle $ is the modified eigenfrequency of the system. Now the new bath operator $\tilde{\Gamma} = \sum_{\mathbf{k} \neq \mathbf{k}_0} g(\mathbf{k}) ( \hat{a}_\mathbf{k}^\dagger \hat{a}_\mathbf{k} -\langle \hat{a}_\mathbf{k}^\dagger \hat{a}_\mathbf{k} \rangle )$ has zero mean and we can substitute it back into Eq. \eqref{gme} and derive the master equation describing the evolution of mode $\mathbf{k}_0$ as
\begin{equation}\label{me_dephase}
\frac{d \hat{\rho}}{dt}=i[\hat{\rho}, \hat{\mathcal{H}}_s] + \kappa_\mathrm{dp} \mathcal{L}_{\hat{n}\hat{n}}(\hat{\rho}),
\end{equation}
where $\hat{\rho}$ is the density matrix describing the mode $\mathbf{k}_0$, $\hat{\mathcal{H}}_s=\omega^{'}_r \hat{a}^\dagger \hat{a}$, $\mathcal{L}_{\hat{n}\hat{n}}(\rho) \equiv 2\hat{n} \hat{\rho} \hat{n}^\dagger - \hat{n}^\dagger \hat{n} \rho - \hat{\rho} \hat{n}^\dagger \hat{n} $ with $\hat{n}=\hat{a}^\dagger \hat{a}$. The parameter $\kappa_\mathrm{dp}$ is a coefficient that characterizes the strength of dephasing
\begin{equation}\label{exchange_integral}
\kappa_\mathrm{dp} = \int_0^\infty  |D(\omega)g(\omega)|^2 n_{\mathrm{th}}[n_{\mathrm{th}}+1]d\omega.
\end{equation}
Note that the modified frequency in the continuum limit reads
\begin{equation}
\omega_r^{'} = \omega_r + \int_0^\infty g(\omega)D(\omega)n_{\mathrm{th}}d\omega.
\end{equation}
At low temperature, this only gives a small correction to the eigenfrequency, hence it can be safely disregarded, i.e., $\omega_r^{'} \approx \omega_r$.  Here the strength of the dephasing rate $\kappa_\mathrm{dp}$ can be comparable to relaxatoin rate of magnons in magnetic insulators and thus has to be considered when manipulating the quantum states of magnons in magnetic insulators at low temperature. An extensive discussion of the dephasing rate $\kappa_{dp}$ can be found in Ref. \cite{YuanArxiv2022}.

\subsection{Magnon-phonon scattering}
Besides the four-magnon scattering process, the magnon-phonon scattering can also result in the dephasing of magnons. Here we consider a one dimensional spin chain along the $z-$axis as an example. The effective Hamiltonian describing the interaction of magnons and phonons \cite{AndreasPRB2014} is $\hat{\mathcal{H}}_\mathrm{int}=\sum_j b_{zz} (\hat{S}_j^z)^2 \hat{\epsilon}_{zz}$, where the strain tensor $\hat{\epsilon}_{zz}=\partial_{z} u_z$ with the atom displacement $u_z=\sum_k (2\rho \omega_kV)^{-1/2}(\hat{b}_k + \hat{b}_k^\dagger)e^{ikz}$. Here $b_{zz}$ is the magnetoelastic coupling coefficient, $\rho$ is the mass density of the magnet, $V$ is the volume, and $\hat{b}_k$ ($\hat{b}_k^\dagger$) is the annihilation (creation) operator of phonon mode with wavevector $k$. The phonon follows the linear dispersion relation $\omega_k = c|k|$ with $c$ being the speed of sound. By substituting the HP transformation of magnons ($S_j^z= S-\hat{a}_j^\dagger \hat{a}_j$) and the quantized phonon mode into $\hat{\mathcal{H}}_\mathrm{int}$, the interaction Hamiltonian can be quantized as
\begin{equation} \label{Hint_real_space}
\hat{\mathcal{H}}_{\mathrm{int}}= \sum_{j} g(\omega_k)\hat{a}^\dagger \hat{a} \left [ \hat{b}_ke^{ikR_j} -\hat{b}_k^\dagger e^{-ikR_j}\right],
\end{equation}
where we have transformed to a discrete space ($z \rightarrow R_j$) and $g(\omega_k)=-2iSb_{zz}k/\sqrt{2\rho \omega_kV}$ is the coupling coefficient of magnons and phonons, which depends on the phonon frequency. Compared with the relaxation process discussed in Sec. \ref{sec_single_mode}, now the magnon density  instead of magnon creation/annihilation operator is coupled to the bath. We will soon see this will result in pure dephasing of magnons without relaxation.

After a Fourier transform of magnon operator $\hat{a}_j =1/\sqrt{N} \sum_k \hat{a}_k e^{ikR_j}$, we simplify the interaction Hamiltonian \eqref{Hint_real_space} as
\begin{equation}
\hat{\mathcal{H}}_{\mathrm{int}}=  \hat{a}^\dagger \hat{a} \sum_k g(\omega_k)(\hat{b}_k - \hat{b}_{k}^\dagger) ,
\end{equation}
where $\hat{a} \equiv \hat{a}_{\mathbf{k}_0}$ is the mode of interest. Here we release the condition of momentum conservation, which may be caused by defects, disorders, grain boundaries or other fluctutations which allows violation of momentum conservation \cite{MichaelJAP2002,SafonovJAP2003}.
Now the system and bath operators read
\begin{equation}
s=\hat{a}^\dagger \hat{a}, \Gamma = \sum_k g(\omega_k)(\hat{b}_k - \hat{b}_{k}^\dagger).
\end{equation}
In the interaction picture, they are rewritten as
\begin{equation}
\tilde{s}=\hat{a}^\dagger \hat{a},\tilde{\Gamma} = \sum_k g(\omega_k)\left (\hat{b}_k e^{i\omega_kt}- \hat{b}_{k}^\dagger e^{-i\omega_kt}  \right).
\end{equation}
By substituting $\tilde{s}$ and $\tilde{\Gamma}$ into the generalized master equation \eqref{gme}, we end up with a similar form of dynamic equation as Eq. \eqref{me_dephase}, but with the dephasing coefficient
\begin{equation}
\kappa_\mathrm{dp} = \frac{8 (Sb_{zz})^2k_BT}{\rho d^2 c^3}.
\end{equation}
Again, the magnitude of $\kappa_{dp}$ can become comparable to the relaxation rate of magnons and a detailed discussion was reported in Ref. \cite{YuanArxiv2022}.
 
\section{Discussions and conclusions}\label{sec_conclusion}
In conclusion, we have presented a master equation approach to address the relaxation and dephasing of magnons in both ferromagnetis and antiferromagnets. Employing this approach, we identified both local and collective dissipation channels of magnons and demonstrated that reduction of noise in spin components below the symmetric limit is a low temperature phenomenon, which makes it different from the classical precession of spins in an elliptical trajectory.  We compared the prediction of magnon relaxation between the quantum formalism and classical dynamic equations. To make them consistent, we have to consider a hybrid bath model, where the magnons on each sublattice of an AFM have their own dissipation channel and also share a common channel mediated by conduction electrons or phonons for example. Up till now, we restricted most of our discussions to the relaxation of uniform precession modes and did not specify the physical origin of the bath explicitly. It would be meaningful to further consider relaxation of magnon modes by rigorously considering the magnon-electron in magnetic metals and magnon-phonon interactions in both metals and insulators. This will help to quantify the temperature and frequency dependence of the relaxation rate $\kappa$ in the master equation.

Besides the master equation approach developed here, the Heisenberg-Langevin equation is another known approach to address the relaxation of quantum states \cite{DFWallsbook}, which has already been utilized to deal with the relaxation of magnons in hybrid magnet-cavity system \cite{Yuanreview2022}. However, the collective dissipation channel through the interaction of magnons is usually not taken into account in this approach. As shown in Sec. \ref{sec_single_mode} and \ref{sec_two_mode}, this collective relaxation channel may be essential to calibrate the squeezing and entanglement of magnons. In the future, it merits further research to study how the interaction of magnons influences the Langevin noise and compare the results presented here with that of the Heisenberg-Langevin equation.

\begin{acknowledgments}
 H.Y.Y acknowledges the European Union's Horizon 2020 research and innovation programme under Marie Sk{\l}odowska-Curie Grant Agreement SPINCAT No. 101018193. A.K. acknowledges financial support from the Spanish Ministry for Science and Innovation -- AEI Grant CEX2018-000805-M (through the ``Maria de Maeztu'' Programme for Units of Excellence in R\&D). R.A.D. is member of the D-ITP consortium that is funded by the Dutch Ministry of Education, Culture and Science (OCW). R.A.D. has received funding from the European Research Council (ERC) under the European Union's Horizon 2020 research and innovation programme (Grant No. 725509).
\end{acknowledgments}

%------------------------------------------------------------------------------%
%\bibliographystyle{apsrev}
\bibliography{master_equation_magnon}

\begin{thebibliography}{55}
\bibitem [{not()}]{noteg1}%
  \BibitemOpen
  \href@noop {} {\enquote {\bibinfo {title} {To avoid the apperance of
  numerical uncertainty in calcuing the normalized coherence funciton for zero
  particle occupation, we plot the unnormalized coherence function in
  \uppercase{F}ig. 3(c).}}\ }\BibitemShut {NoStop}%
\end{thebibliography}
\begin{thebibliography}{99}
\bibitem{noteg1} To avoid the apperance of numerical uncertainty in calcuing the normalized coherence funciton for zero particle occupation, we plot the unnormalized coherence function in Fig. \ref{fig3}(c).
\end{thebibliography}
\begin{comment}

\end{comment}
\end{document}